\documentstyle[prb,aps,multicol,psfig]{revtex}
\tighten
\begin{document}
\title{Polaronic optical absorption of the Holstein and  
Su-Schrieffer-Heeger models}
\author{M. Capone and W.Stephan\cite{newaddress}}
\address{Istituto di Fisica della Materia and
Dipartimento di Fisica, Universit\`a di Roma ``La Sapienza'',\\
Piazzale A. Moro 2, Roma, Italy 00185}
\maketitle 

\begin{abstract}
The optical conductivity of a polaronic charge carrier in the 
intermediate and strong-coupling regimes is calculated for
a tight-binding electron using exact diagonalization.
Two different simple models of the electron-phonon coupling
are considered:
the Holstein model, with coupling of the lattice displacement
to the local electron density, and the Su-Schrieffer-Heeger
type of coupling arising from modulation of the electronic
overlap integral.  The two models
are shown to exhibit a similar polaronic
absorption band.  However,
the Su-Schrieefer-Heeger form of coupling gives rise to
a second strong absorption band not present in the
Holstein model.  This additional feature corresponds
to transitions from the ground state of bonding 
character on a shortened bond to excited states of anti-bonding 
character on the same shortened bond.
\end{abstract}


\begin{multicols}{2}

\section{Introduction}

Electrons may acquire polaronic character when they strongly 
interact with phonons in such a way that the effect of the 
interaction is not merely a perturbative correction to the 
free electron. More specifically, if the electron-phonon 
($e$-$ph$) interaction is sufficiently strong a cross-over occurs 
from the quasi-free electron limit in which the $e$-$ph$ interaction 
weakly renormalizes the free-electron properties, to the polaronic 
state in which the electron is strongly bound to a potential well 
generated by the lattice distortion induced by the electron itself.

The possible relevance of polaronic phenomena to the high $T_c$
superconductors suggested by the observation of polaronic features 
in the optical response of lightly doped cuprates              
\cite{polaronexper,polaronexper2} has generated renewed 
interest in the theoretical study of the
conditions for and consequences of polaron formation.
The strong-coupling nature of the polaron state severely limits
the possibilities for reliable analytic
approaches, in particular in the intermediate coupling region near 
the cross-over. For example, while quite accurate variational wave 
functions may be formulated\cite{Sergio}, it is not feasible to 
calculate dynamical properties within this approach.
The numerical exact diagonalization of small clusters on the other 
hand, allows the exact calculation of both static and dynamic 
properties with the only limitations being the finite memory 
available on the computer. We therefore use exact diagonalization 
to determine the optical conductivity for small one-dimensional 
clusters.  Since the polaron becomes well localized for 
sufficiently strong coupling, finite-size effects may be controlled 
to a large extent by appropriate choice of the model parameters.

In this paper we study the optical response of the Holstein model 
and the Su-Schrieffer-Heeger (SSH) model. These models are the most 
used for electrons interacting with phonons via a short range 
$e$-$ph$ interaction; we study the analogies and the differences 
between the optical absorption features for the two models.

The Holstein model is described by the Hamiltonian
\begin{equation}
\label{holham}
{\cal H}=-t\sum_{\langle ij\rangle}
c^\dagger_i c_j 
+\omega_0 \sum_i a^\dagger_i a_i
+ g\sum_i 
c^\dagger_i c_i \left( a_i+a^\dagger_i \right)
\end{equation}
and the Su-Schrieffer-Heeger model by the Hamiltonian
\begin{eqnarray}
\label{sshham}
{\cal H} & = & -t\sum_{\langle ij\rangle}
c^\dagger_i c_j 
+\omega_0 \sum_i a^\dagger_i a_i \nonumber \\
& + & g\sum_i 
\left[ \left( c^\dagger_i c_{i+1} +
 c^\dagger_{i+1}c_i \right)
\left( a^\dagger_{i+1}+a_{i+1}-a^\dagger_i-a_i \right) \right].
\end{eqnarray}

We use units such that $\hbar = c = a = 1$, where $a$ is the 
lattice spacing. In both models the free electron part (first term) 
corresponds to hopping to a  nearest neighbour site with hopping 
integral $t$, which we will take as our unit of energy. Since we 
restrict ourselves to the single electron case we will suppress 
electron spin indices. The second term decribes free Einstein 
dispersionless phonons ($\omega(q) = \omega_0$) on each site.
The $e$-$ph$ term introduces a coupling between the phonon field 
and the local electron density (Holstein model) or the covalent 
bond operator $c^{\dagger}_{i}c_{j}$ (SSH model).
Note that this model differs from the conventional SSH model in having
local Einstein phonon modes rather than acoustic phonons.
To compare and contrast the optical absorption of the two models
we avoided unnecessary differences between them, in order to
identify the differences deriving from the different $e$-$ph$ 
coupling.

The Holstein $e$-$ph$ coupling arises from the dependence of the 
atomic potential energy on the ionic displacement. This coupling 
is relevant in physical systems in which the screening is very 
weak, as is believed to be the case in the high $T_c$ 
superconductors with small doping\cite{zeyher}. On the other hand, 
the SSH coupling arises if one assumes that the hopping integral 
depends on the distance between the ions. Despite the fact that 
the physical origin of the two terms is different we will use the 
same notation for the bare coupling constant ($g$) for both models, 
in order to underline the general character of the physical 
processes under consideration.
One may define two independent dimensionless couplings, $\alpha = 
g/\omega_0$ and $\lambda = g^2/(2td\omega_0)$ where $d$ is the 
dimensionality. In a previous paper\cite{criterio} we studied the 
conditions for the formation of small polarons for both models, 
finding that for the Holstein model the carriers have polaronic 
character if $\lambda > 1$ if the phonon frequency $\omega_0$ is 
smaller than $t$ (adiabatic regime), while the criterion is
$\alpha > 1$ if the phonon frequency is larger than the hopping 
integral (antiadiabatic regime).  Note that since polaron formation 
is not a phase transition, but rather a  cross-over, there is some 
ambiguity in the definition of the exact numerical value of the 
critical value for polaron formation.
For the SSH model the relevant parameter is $\lambda$ regardless 
of the value of the adiabatic ratio $\omega_0/t$. $\lambda = 0.2$ 
is the critical value for a significant polaronicity of the ground 
state of this model.

One important and direct consequence of the different nature of the 
electron-phonon coupling in the two models is obvious in the 
strong-coupling  limit: the local character of the Holstein 
coupling drives the system towards polaron states localized on a 
single site, while the covalent coupling in the SSH model leads 
the system to localize the electron on a bond, forming an even 
symmetry state localized on two neighbouring sites. We shall see 
that this difference between ``site'' and ``bond'' polarons
manifests itself in very obvious differences in the optical 
absorption spectra of the two models.

\section{Formalism}
\label{sec:formal}

The real part of the conductivity for a one-dimensional 
tight-binding model at zero temperature may be expressed in 
terms of the Kubo formula
\begin{equation}
\sigma(\omega) = D\delta(\omega) + 
\Im \langle 0\vert J^\dagger {1\over \omega-H+E_0-i\delta}J\vert 
0\rangle
\label{kubo},
\end{equation}
where $J$ is the current operator. 

The coefficient of the zero-frequency delta function contribution
$D$ is usually called the Drude weight: it is given by 
\begin{equation}
\label{drude}
D = -{\pi e^2\over 2}\langle H_t \rangle - 
\sum_{n\neq 0}{\vert\langle\phi_0
\vert J\vert\phi_n\rangle\vert^2\over E_n - E_0}.
\end{equation}
If the Drude weight $D$ is non-zero the system is a perfect
conductor\cite{kohn}; this will generally be the case in 
such models with periodic boundary conditions (PBC) and no 
disorder at zero temperature.

In the Holstein model $J$ is 
\begin{equation}
\label{currenthol}
J_H = iet\sum_{i} \left(c^\dagger_{i+1}c_{i}-c^\dagger_{i}c_{i+1} 
\right)
\end{equation}
while for the SSH model it is
\begin{eqnarray}
\label{currentssh}
& &J_{SSH} = ie\sum_{i\sigma}\left[t - g(a^{\dagger}_{i+1} + a_{i+1} - 
a^{\dagger}_i
 - a_i)\right]\times \nonumber \\
& & \left(c^{\dagger}_{i+1}c_{i} - c^{\dagger}_{i}c_{i+1}\right).
\end{eqnarray}
The fact that $J_{SSH}$ contains an explicit coupling
to the phonon degrees of freedom is physically simple to understand:
in this case the bond-length is modified by the lattice distortion,
so that the change in electric dipole moment associated with the
hopping of an electron is modified proportionally.  Here we have
neglected in both cases the direct coupling of the electric field 
to the ions which is of order $\sqrt{m/M}$, where $m$ and $M$ are 
the 
electron and ion masses.  We are interested only in the features 
resulting from the $e$-$ph$ coupling, not in the direct excitation 
of the bare phonons. Equations (\ref{kubo}), (\ref{drude}), 
(\ref{currenthol}) and (\ref{currentssh}) may be derived following 
the standard approach used for example in the case of the Hubbard 
model\cite{Shastry}.

The use of the Lanczos algorithm to evaluate correlation functions
such as (\ref{kubo}) is well established \cite{dagotto}.
In the present models one must introduce a cutoff in the occupation
of the phonon modes in order to represent the Hamiltonian as a finite
matrix \cite{ranninger}.  We have checked that the results presented
here are not significantly affected by this cutoff by comparing 
results for different cutoffs.  The finite-frequency part
of (\ref{kubo}) has previously been studied by Alexandrov et al.
\cite{akr} for the Holstein model.

We have used both periodic and open boundary conditions for our 
numerical calculations. With open boundary conditions (OBC)
the system is like a molecule; in this case free acceleration 
is impossible and the Drude weight $D$ is always zero. With PBC 
$D$ may be determined either by studying the dependence of 
the ground state energy on an adiabatic change of boundary 
condition, equivalent to a static uniform vector potential 
\cite{kohn}, or by making use of the f-sumrule which relates the 
integrated conductivity to a ground state expectation value.
For the Holstein model the latter is
\begin{equation}
\label{sumhol}
\int_0^{\infty}\sigma(\omega)d\omega = 
-{\pi e^2\over 2}\langle H_t\rangle,
\end{equation}
while, given the explicit form for the current for the SSH model 
Eq.(\ref{currentssh}) the sumrule for the SSH model is given by
\begin{equation}
\label{sumssh}
\int_0^{\infty}\sigma(\omega)d\omega = -{\pi e^2\over 2}\langle H_t + 
H_{e-ph}\rangle.
\end{equation}
The $e$-$ph$ coupling term appears in the sumrule for the SSH model 
in the same way the hopping term does: this is a direct 
consequence of the origin of this term, arising from a modulation
of the hopping integral.  We have made use of the sumrule
to determine $D$.
  
\section{The optical excitation of small polarons: simple limits}

The optical excitation of a small polaron for the Holstein model 
has been studied by Emin\cite{emin} by means of general arguments 
in the adiabatic limit and calculated by means of exact 
diagonalization by Alexandrov et al. \cite{akr} for the more 
general case.
The physical origin of the optical absorption of a small polaron 
can be easily described in the adiabatic limit $\omega_0 = 0$ 
invoking the Franck-Condon principle.
The ground state is given by an electron localized on a single 
site, which is strongly displaced from its equilibrium position, 
while all the other sites are not displaced. The electron can be 
excited to a neighbouring site without changing the lattice 
configuration by the application of the current operator.
The difference in energy between the two states is the lowering 
of the electronic energy associated with the small polaron 
formation $2E_b$, where $E_b = 2\lambda t$ is the small polaron 
binding energy.

The physical mechanism we have described is not peculiar to the 
extreme adiabatic limit $\omega_0 = 0$, and is not strongly 
affected by the introduction of the lattice dynamics via a finite 
value of the phonon energy $\omega_0$.  Note that the current 
operator (\ref{currenthol}) acts only on the electronic degrees 
of freedom. Hence the current operator connects only states having 
the same lattice configuration, or at least having a non-zero 
overlap as far as the phononic state is concerned. Thus the 
physical picture we introduced for the extreme adiabatic limit
can be extended to finite frequencies.
 
The small polaron optical conductivity is then characterized by 
an absorption band centered at $2E_b$,  well described by Reik's 
formula\cite{reik} 
\begin{equation}
\label{reik}
\sigma(\omega) = {t^2\over 2E_bT}
{1-e^{-\beta\omega} \over \omega}exp\biggl[
-{(\omega-2E_b)^2 \over 8E_bT}\biggr];
\end{equation}
The width of the band is proportional to $\sqrt{E_b T}/t$, which, 
in the small $T$ limit reduces to $\sqrt{E_b\omega_0}/t = g/t$, 
while the maximum of the intensity is proportional to $t^2/E_b T$, 
which reduces to $t^2/g^2$ in the small $T$ limit.

The SSH model optical conductivity can also be studied starting 
from adiabatic arguments invoking the Franck-Condon principle, 
but now taking into account the bond nature of the polaronic state.
The ground state is characterized by a short bond on which the 
electron is localized. The short bond is generated by the shift 
of two neighbouring sites towards one another by equal amounts. 
The electronic ground state is the even combination of the two 
local states.
 
The optical absorption can happen in two different channels: a 
``Holstein-like'' one in which the electron is excited onto a 
different bond that is not shortened, and a local channel in 
which the electron is excited from the even symmetry ground 
state into the local odd symmetry state on the short bond.
The first kind of excitation is analogous to the excitation
of the Holstein polaron and is expected to generate a band 
similar to the one previously described,
centered at $2E_b$ (even if in this case $E_b$ is not simply 
given by the Lang-Firsov result $g^2/\omega_0$). 
The local excitation energy is given by the difference 
in electronic energy between the even parity ground state and 
the odd parity state, keeping the lattice configuration fixed.
While the ``Holstein-like'' excitation is characterized by the 
fact that the electron is excited from a state in which it gains 
an energy $2E_b$ from the local distortion to a state in which 
the electron energy is not affected by the lattice configuration, 
the ``local'' transition carries the electron from a state in 
which the distortion lowers the energy by an amount $2E_b$ to a 
state in which the electron energy is raised by the same amount 
$2E_b$. Hence the energy difference involved in the optical 
transition is $4E_b$. If we introduce a finite phonon frequency 
this absorption peak broadens into a ``band'' exhibiting phonon 
features separated by the typical phonon frequency $\omega_0$.  

For the special case of only two sites the SSH model can be 
analytically solved for arbitrary $\omega_0$ by means of a 
modified Lang-Firsov transormation that acts on the bond 
variable. After performing the modified Lang-Firsov 
transformation we obtain for the two site cluster
\begin{equation}
\label{ssh2}
\sigma_{2s}(\omega) = {\pi e^2\over 4}\sum_{n=0}^{\infty}(2t + 
n\omega_0)e^{-8\alpha^2}
{(8\alpha^2)^n \over n!}\delta(\omega - (2t + n\omega_0)).
\end{equation}

The conductivity of the two site cluster given by (\ref{ssh2})
consists of a succession of Dirac delta functions at frequencies 
separated by the phonon frequency $\omega_0$. 
Once the Dirac delta functions are substituted by Lorentzians 
the analytical formula (\ref{ssh2}) gives the same result as the 
numerical calculations: a single absorption band centered at 
$\omega = 2t + 8g^2/\omega_0$, with width proportional to $g/t$ 
and intensity  proportional to $t/g + 4g/\omega_0$.
Of course the other ``Holstein-like'' absorption feature
which we argued should be present due to the transfer of the 
electron from the shortened bond to a neighboring undistorted 
one cannot occur for a two site system where there is only one 
bond. This implies that the sumrule (\ref{sumssh}) which is given by
\begin{equation}
\int_0^{\infty}\sigma(\omega)d\omega = {\pi e^2\over 2}\bigg(t + 
{4g^2\over\omega_0}\bigg)
\end{equation}
is exhausted by this feature. This is consistent with the width and 
intensity described above.

We expect the physics of the polaronic absorption for the SSH model 
to be much more dependent on the adiabatic ratio than is the case 
for the Holstein model. The dependence of the current operator on 
the phonon operators makes it possible to have an optical transition 
which does not leave the lattice configuration unaltered, expecially 
if the phonon frequency is comparable or even greater than the 
hopping integral.  This point will be discussed further in light 
of the numerical results.

\section{Exact diagonalization results}

In Fig. 1 we show the optical conductivity for the Holstein and
SSH models for different lattice sizes with open boundary 
conditions, for a single particle with a phonon frequency 
$\omega_0 = 0.2t$. The lattice sizes range from two to four sites 
from top to bottom, with the left column presenting results for 
the Holstein model and the right column analogous results for 
the SSH model. We consider only one-dimensional systems, but
due to the local nature of the polarons we are considering and 
the arguments we previously introduced we do not expect the 
dimensionality to play a crucial role.
Therefore, even if the numerical calculations refer to the 
one-dimensional case, the physical ideas we are describing are
more general.

\begin{figure}
{\hbox{\psfig{figure=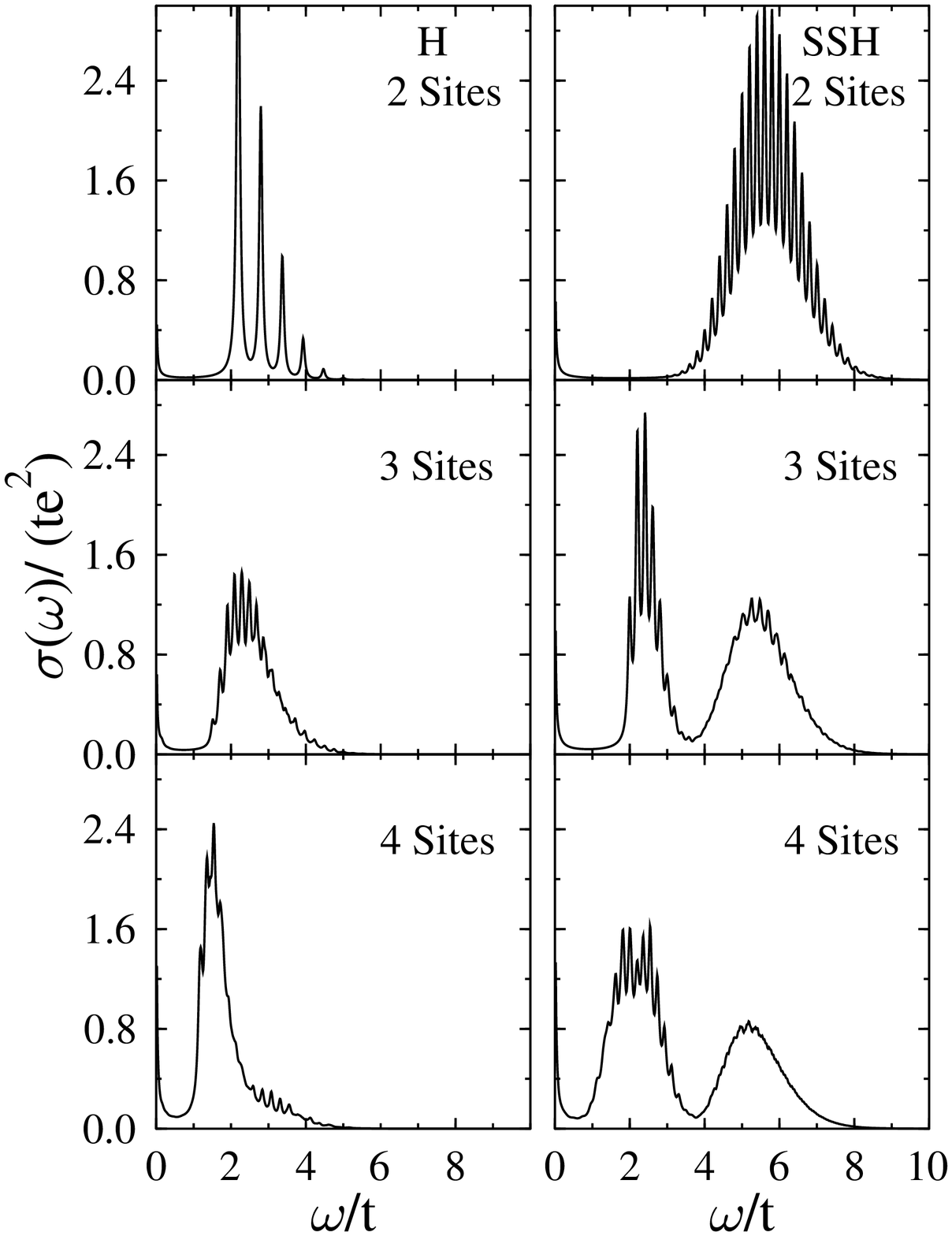,width=8.0cm,angle=0}}}
\vspace{-0.5truecm}
\small{FIG. 1: Finite frequency optical conductivity
for the Holstein model
(H, left column) and the SSH model (SSH, right column) for different
lattice sizes: from top to bottom 2, 3 and 4 sites.}\\
\end{figure}

We have chosen the coupling for the two different models in order 
to have the low energy feature for the SSH model centered at the 
same frequency as the feature in the Holstein model. 
For the Holstein model this coupling is intermediate, 
which can be seen both from the asymmetry of the absorption band
and the significant size dependence of the spectrum \cite{akr}.
With a further increase in the coupling strength this absorption
band becomes very similar to that expected from simple analytic 
approaches \cite{emin,reik} as was previously found in ref.
\onlinecite{akr}. For the SSH model on the other hand, we are 
already deeper in the polaronic regime. This is clear from the 
fact that the low energy feature for the SSH model does not 
change significantly from 3 to 4 sites while the one in the 
Holstein case changes more noticeably. If one increases the 
coupling strength further however, there is the risk of obtaining 
unphysical results for the SSH model. Due to the linearized form 
of the $e$-$ph$ coupling, when the displacement becomes 
sufficiently large the effective hopping matrix element may 
pass through zero and change sign. In a more complete treatment 
of the $e$-$ph$ coupling this would be counteracted by non-linear 
terms. In the present work we avoid the unphysical region of 
this simple version of the model by restricting the 
coupling strength.
The numerical calculations have been performed with a maximum 
allowed number of phonons $n_{max} = 50$ for the two and three 
site calculations and $n_{max} = 20$ for the four site calculations. 

While the results for the Holstein model do not depend strongly 
on the number of sites, the SSH two-site model has a very different 
behaviour compared to the larger systems.
The two-site SSH model, as we have anticipated in the previous 
section shows just a single feature centered at $\omega = 2t + 
8g^2/\omega_0$, associated with a local transition from the even 
parity ground state to  odd parity excited states.
Increasing the number of sites a lower energy feature appears 
centered at half the centre of the high energy ``local'' feature. 
It is generated by the excitation of the electron from one bond to 
a neighbouring one. It is worth  noting that the two features do 
not change significantly going from 3 sites to 4 sites, despite 
the fact that the model has a non-local $e$-$ph$ coupling.  

To summarise the results up to this point: the most noticeable 
feature of the optical conductivity of a single SSH polaron in the 
(quasi-)adiabatic limit is the presence of two optical absorption 
bands generated by different optical excitation processes, one 
corresponding to the feature found in the Holstein model, and 
the other at twice the energy corresponding to a local excitation 
on the distorted bond.

Further information about the different way in which the polaron 
excitation occurs in the two models  can be extracted from an 
analysis of the optical spectra as functions of the $e$-$ph$ 
coupling. This may be most clearly seen by examining the 
various contributions to the sumrule.  As a consequence of 
the self-trapping, the electronic kinetic energy is strongly 
suppressed in the strong coupling limit; Eq.(\ref{sumhol})
implies that the total weight of optical excitations decreases 
with increasing coupling for the Holstein model.
On the other hand the optical sumrule for the SSH model 
Eq.(\ref{sumssh}) also involves the $e$-$ph$ term, which increases 
with increasing coupling constant. 

As mentioned in Section \ref{sec:formal}, the Lanczos algorithm 
allows us to calculate exactly the finite-frequency part of the 
conductivity, which coincides with the total conductivity for OBC.
Using PBC, the additional information from the sumrule allows the 
Drude weight to be determined.
In Fig. 2 we show for the Holstein model the total sumrule,
the Drude weight and the incoherent integrated weight for a 
four-site cluster with PBC and a phonon frequency $\omega_0 = 0.2t$.
The total sumrule sharply decreases as soon as $\lambda \sim 1$; 
this sharp decrease is driven by the fall of the Drude weight, 
which rapidly approaches zero. Note further that even if, 
for $\lambda > 1$, the polaronic peak given by Reik's formula 
appears besides the Drude weight, its weight also decreases as 
$t/g$ as $g$ increases in the strong-coupling regime.

\begin{figure}
{\hbox{\psfig{figure=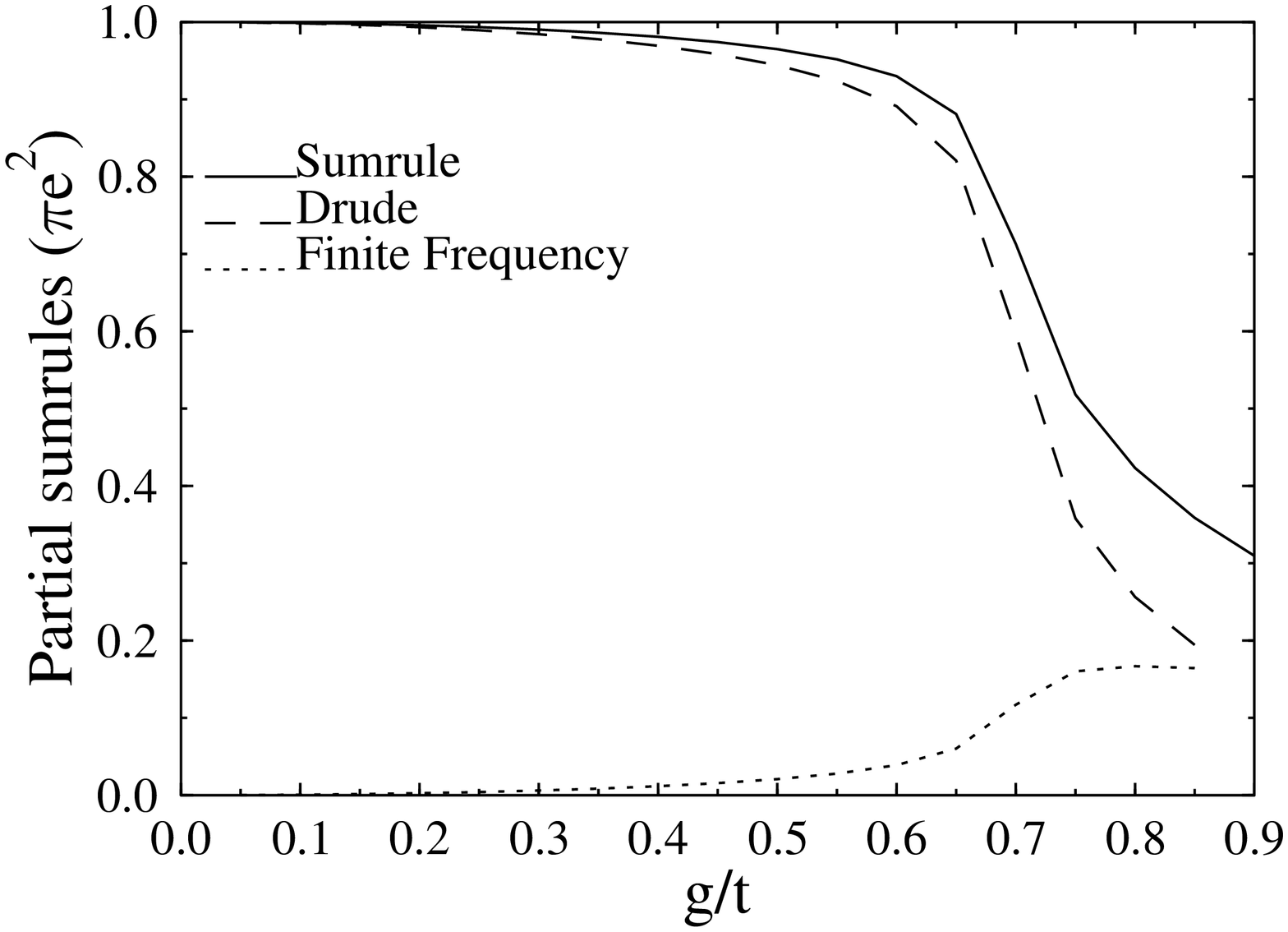,width=5.5cm,angle=-90}}}
\vspace{-1.4truecm}
\small{FIG. 2: Spectral weights for the Holstein 
model for a four site cluster with PBC. The solid line is the total 
sumrule Eq.(\protect\ref{sumhol}); the dashed line is the Drude weight; 
the dotted line is the integrated weight of the finite frequency 
polaronic absorption feature.}\\
\end{figure}

\vspace{-1.7truecm}
\begin{figure}
{\hbox{\psfig{figure=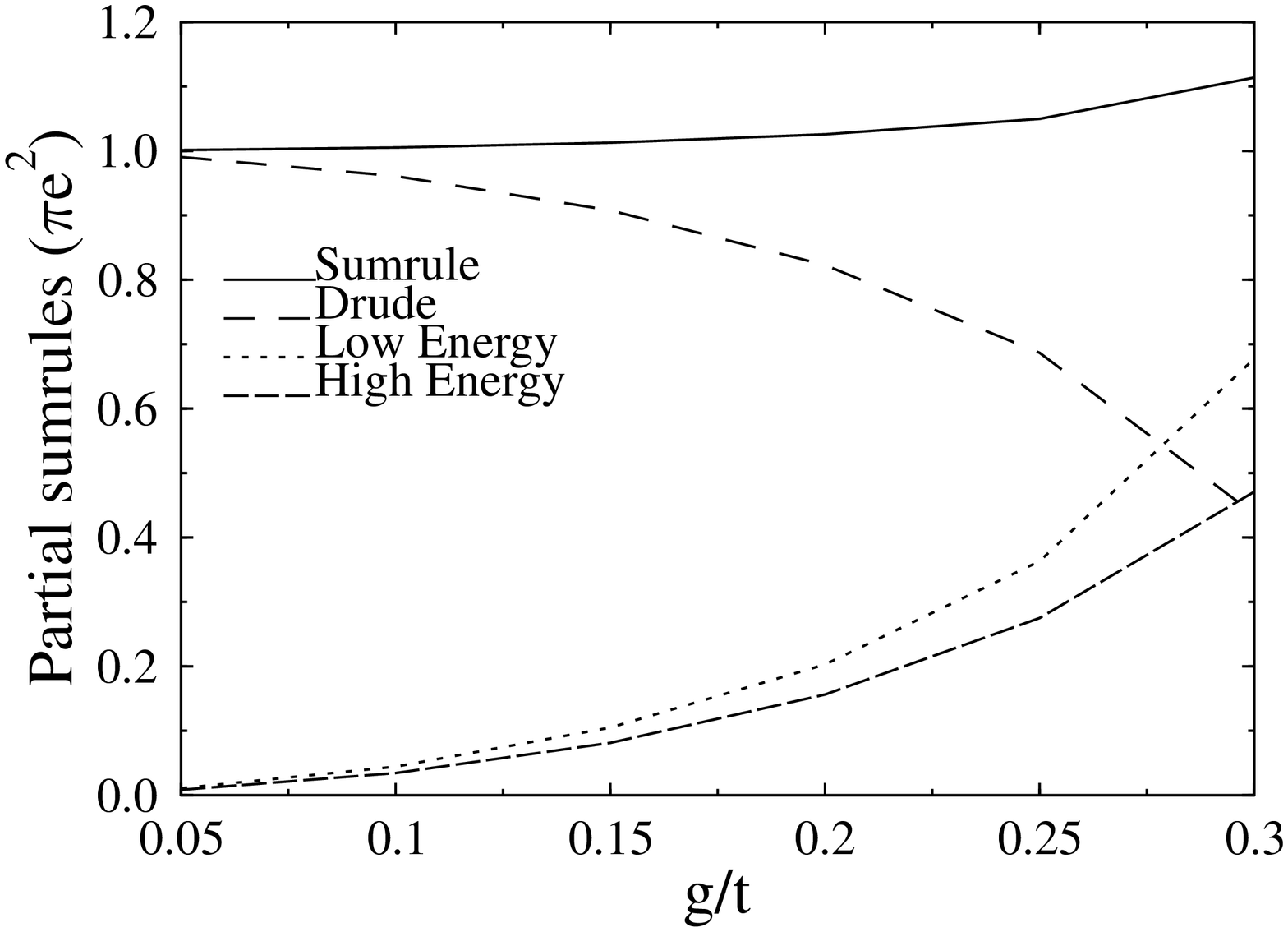,width=5.5cm,angle=-90}}}
\vspace{-1.4truecm}
\small{FIG. 3: Spectral weights for the SSH 
model for a four site cluster with PBC. The solid line is the total 
sumrule Eq.(\protect\ref{sumssh}); the dashed line is the Drude weight; 
the dotted line is the integrated weight of the low energy 
``Holstein-like'' polaronic absorption feature; the dot-dashed line
is the high energy local bonding-antibonding transition.}\\
\end{figure}

In Fig. 3 we present similar information for the SSH model,
showing the results for a four-site cluster with PBC for a phonon 
frequency $\omega_0 = 0.2t$.
The total sumrule monotonically increases with the coupling, which 
indicates that the increase of the $e$-$ph$ term overcompensates 
the decrease of the hopping term.
A detailed analysis of the different contributions to the optical 
sumrule shows that the Drude weight rapidly decreases for this kind
of $e$-$ph$ coupling as well as for the Holstein one, whereas the 
polaronic structures increase their total weight as the coupling 
increases.

We have separately integrated the optical conductivity for the 
two different optical absorption bands characteristic of the SSH 
case, obtaining in both cases an increase in total optical weight
with coupling. As far as the high-energy, local feature is concerned 
this result is consistent with Eq. (\ref{ssh2}), which predicts an 
increasing value for the intensity of the optical absorption as a 
function of the $e$-$ph$ coupling.
The low-energy feature for the SSH model on the other hand, 
which  we attributed to the same kind of optical excitation that 
generates the absorption band for the Holstein model,
also shows an increase of total weight with increasing $g$, 
whereas the Holstein structure has a decreasing weight as the 
coupling increases. This difference does not undermine the 
similarity between the two features, but simply underlines the 
nature of the $e$-$ph$ coupling term for the SSH model:  
the dipole moment associated with this type of transition is 
an increasing function of the coupling, so that quite naturally 
the absorption is expected to increase with coupling.

A further aspect to be considered is the dependence of the 
conductivity on the phonon frequency. As we already stated, 
the independence of the current operator for the Holstein model 
on the phonon operators is responsible for the ``survival'' of the 
adiabatic small polaron excitation process with increasing phonon 
frequency. For the SSH model this argument does not work, so we 
expect that the physical picture we have drawn using the adiabatic 
approximation will not hold for a sufficiently large value of the 
adiabatic ratio $\omega_0/t$. The dependence of the current operator 
on the difference of the phonon displacements makes it possible 
to consider an optical transition that modifies, even strongly, 
if $\omega_0/t$ is sizeable the lattice configuration. The short 
bond can be enlarged as the electron is excited to a neighbouring 
bond and the adiabatic picture can break down.

\begin{figure}
{\hbox{\psfig{figure=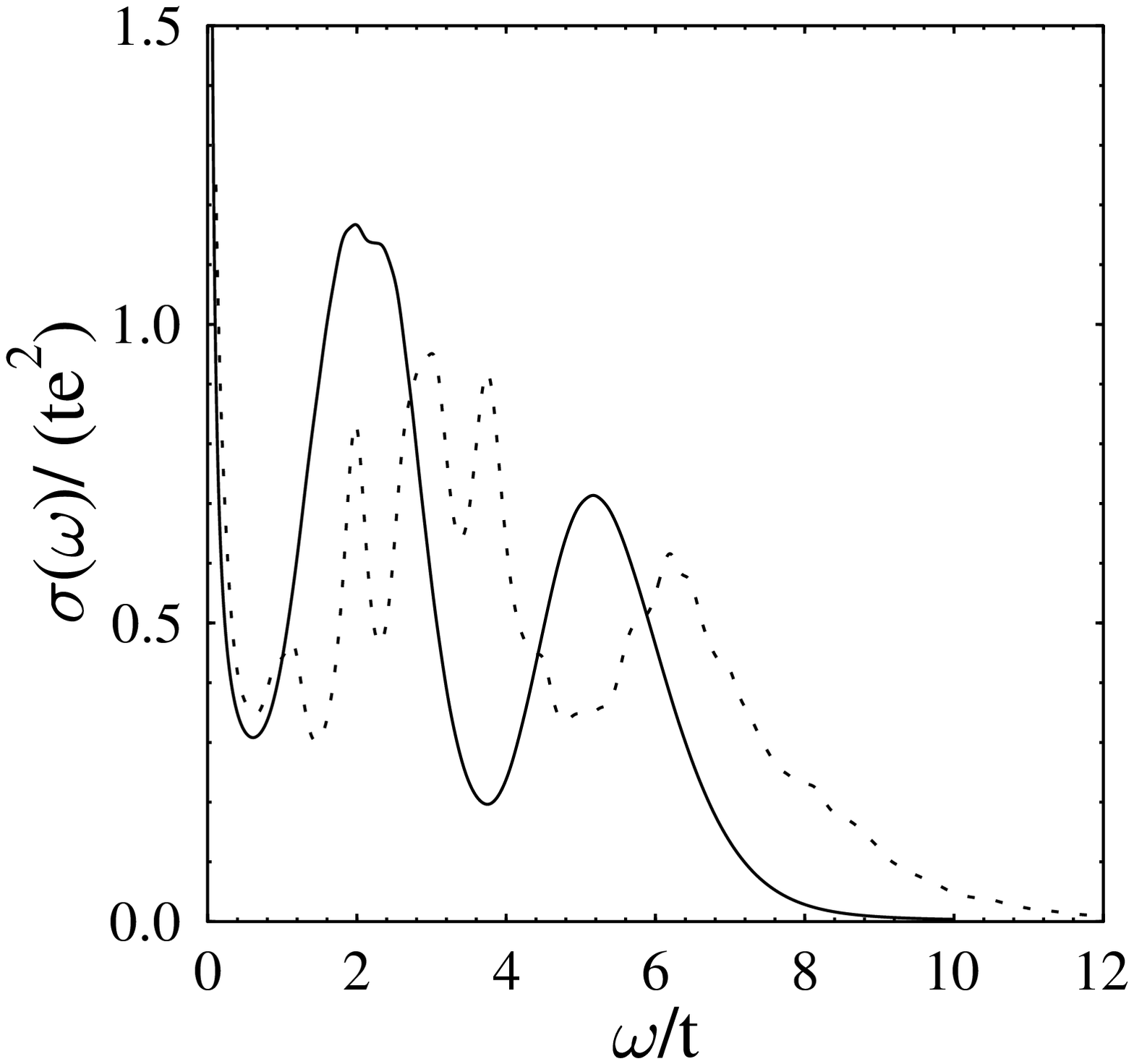,width=7.5cm,angle=0}}}
\vspace{-0.5truecm}
\small{FIG. 4: Finite frequency optical conductivity of the 
SSH model for a four site cluster using $\lambda =0.225$. 
The solid line is for $\omega_0/t =0.2$ and the dashed line
is for $\omega_0/t =0.8$.}\\
\end{figure}

In Fig. 4 we study the effect of increasing phonon frequency 
on the optical conductivity for a four-site OBC cluster keeping 
$\lambda$ constant. The value of the coupling ($\lambda = 0.225$) 
is large enough to show clear polaronic features in the adiabatic 
regime. The two optical features are very evident for the smaller 
of the two frequencies shown $\omega_0/t=0.2$, whereas for the 
larger phonon frequency $\omega_0/t=0.8$ the absorption bands are 
broadened and overlap significantly. Larger phonon frequency also
results in a more evident phonon structure, which contributes to 
hide the two features. For an adiabatic ratio $\omega_0/t = 1$ 
there is almost no sign of two distinct optical structures.
Note that for larger phonon frequency size effects are also more
important:  for $\omega_0/t=0.8$ the difference between the high 
energy region of the 4-site result shown and the 2-site case is 
much larger than is the case for the results for $\omega_0/t=0.2$ 
in Fig. 1.

\section{Conclusions}

In this paper we have studied the optical conductivity
of the Holstein and SSH models in the intermediate and 
strong-coupling polaronic regime.
The optical absorption of a polaron arising from $e$-$ph$ 
coupling of the SSH type has been shown to exhibit marked 
differences from the well-known Holstein polaron.  
These differences have been shown to be a consequence of the 
nature of the polarons in the two models, which is in turn
a consequence of the different way the electrons are coupled
to the phonons. Whereas the Holstein small polaron
consists of an electron bound to a single distorted site,
the SSH polaron may be described as an electron localized on
a shortened bond.  

The absorption in the Holstein model 
is due to processes where the electron is excited from 
the distorted site to a neighbouring
undistorted site. A feature in the optical conductivity
centered at a frequency $\omega = 2E_p$, i.e. twice the
polaron binding energy, is associated with this kind of 
process. In the SSH model corresponding processes exist
where the electron is transferred from the distorted bond
to a neighbouring undistorted bond, leading to a very 
similar absorption feature.
On the other hand, a different channel for the polaron 
excitation in the SSH model is available. The ground state is
an even parity, bonding state localized on a shortened bond.
Due to the existence of local excited states of anti-bonding 
(odd) symmetry on the ``short'' bond, there is an additional 
strong absorption feature at twice the energy of the familiar 
``Holstein-like'' absorption. The higher energy of this feature 
may be understood to arise from the anti-bonding nature of 
these final states with respect to the shortened bond.
This anti-bonding character leads to the raising of the energy 
of the state by the same amount by which the polaronic ground 
state energy is lowered, whereas the ``Holstein-like'' transition 
occurs from a low energy state to a zero energy state as far as 
the electron-phonon interaction energy is concerned.
Although all numerical calculations were performed for 
one-dimensional systems, due to the local nature of the physics 
we have described, the dimensionality is not expected to play a 
crucial role and similar features would be expected also
in higher dimensional systems.

\acknowledgments 
We acknowledge fruitful discussions with 
Prof. P. Calvani, Prof. C. Castellani, Dr. S. Ciuchi, Prof. C. Di Castro
and Dr. M.Grilli. This work was partly supported by the
Istituto Nazionale di Fisica della Materia (INFM).

\end{multicols}

\end{document}